\documentstyle[aps,prl]{revtex}

\begin{document}

\input epsf.sty
\twocolumn[\hsize\textwidth\columnwidth\hsize\csname %
@twocolumnfalse\endcsname

\draft

\widetext

\title{Magnetic frustrations in the face centered cubic
  antiferromagnet NiS$_{2}$}

\author
{
M. Matsuura$^{1,3}$, Y. Endoh$^{2,3}$,
H. Hiraka$^{2}$, K. Yamada$^{4}$, K. Hirota$^{1}$, A.\ S.\
Mishchenko$^{5,6}$,
N. Nagaosa$^{5,7}$ and I.\ V.\ Solovyev$^{8,9}$ }

\address
{
$^{1}$Department of Physics, Tohoku University, Sendai 980-8578, Japan\\
$^{2}$Institute for Materials Research, Tohoku University, Katahira,
Aoba-ku, Sendai 980-8577, Japan\\
$^{3}$Core Research for Evolutional Science and Technology, JST, Kawaguchi
332-0012, Japan\\
$^{4}$Institute of Chemical Research, Gokasho, Uji, Kyoto 611-0011, Japan\\
$^{5}$Correlated Electron Research Center, Tsukuba, Ibaraki 305-0046,
Japan\\ $^{6}$Russian Research Centre ``Kurchatov Institute'', 123182
Moskow, Russia\\
$^{7}$Department of Applied Physics, University of Tokyo, Bunkyo-ku, Tokyo
113-8656, Japan\\
$^{8}$JRCAT-Angstrom Technology Partnership,
Tsukuba, Ibaraki 305-0046, Japan\\
$^{9}$Institute of Metal Physics, Russian Academy of Sciences, Ekaterinburg
GSP-170, Russia
}

\date{\today}
\maketitle
\begin{abstract}
Neutron scattering experiment on NiS$_2$ single crystal
revealed a honeycomb pattern of the intensity distribution
in reciprocal lattice space
(continuous-line structure along the $fcc$ zone boundary )
providing the first direct evidence for nearly frustrated
antiferromagnetism (AF) on the face centered cubic ($fcc$) lattice.
A small but finite lattice distortion below 30.9 K lifts the degeneracy of
the magnetic ground state due to the frustration and eventually result in
the coexistence of the type I and the type II AF LROs, which are mutually
incompatible in the $fcc$ symmetry at higher temperatures.
\end{abstract}

\pacs{ 75.25.+z, 75.40.Gb, 75.50.Ee}

\phantom{.}

]

\narrowtext

In spite of extensive theoretical and experimental explorations \cite{Ra01}
of frustrated magnetism,
search for such new examples is still very active.
This class of materials does provide many new and unexpected phenomena
including the high temperature superconductivity.
Therefore the research and development for such materials constantly
brings new challenges for theoretical interpretation.
These materials often show the robust magnetic frustration phenomenon.
A prototype is a Kagom\'e antiferromagnet, 
where the magnetic frustration is
caused by solely rigid geometrical spin structure
with a nearest neighbor (NN) antiferromagnetic (AF) interaction
on a 2-dimensional (2D) triangular lattice \cite{Ra01}.
Extensive studies of this model have revealed that 
suppression of a long range order (LRO) 
in geometrically frustrated compounds is directly affected
by strong quantum fluctuations.
Another example is the 3D antiferromagnet with corner-sharing tetrahedra
such as transition metal pyrochlores and spinels \cite{Ra01}.
On the other hand, one can introduce a
fragile frustration phenomenon which
is subject to both geometrical spin
structure and a subtle balance among various magnetic interactions.
Such a subtle balance is easily broken
by structural phase transition (SPT) as observed experimentally
\cite{Lee00,Melzi00}.

In order to study the spin dynamics of these systems, the
momentum and energy dependence of the generalized spin susceptibility
$\chi^{\prime\prime}({\bf q}, \omega)$ is indispensable.
However in most cases, the lack of large enough single crystal
prohibits
the detailed studies on the momentum dependence,
and often the parameters in the theoretical model have been set
rather arbitrarily. It is emphasized here that the geometrical structure
only is not enough to guarantee the frustration and degeneracy.
In this paper we report the neutron scattering experiment on the
NiS$_2$ single crystal with $fcc$ lattice structure.
The detailed analysis of the
scattering intensity provides the first direct evidence that the
system is very near the phase boundary between
the two types of antiferromagnetic long range order (AF LRO). 
This quasi-degeneracy is lifted by the SPT at low temperatures.

NiS$_{2}$ possesses anomalies in its thermodynamic properties
\cite{rf:yao96,Yao97,rf:00matsuura}
and an unusual coexistence of type I and 
type II antiferromagnetic (AF) LROs \cite{rf:hasting70,rf:78kikuchi}
which are incompatible from symmetry considerations
for the undistorted $fcc$ lattice.
It shows magnetic phase transitions at low temperatures with
persistent short range magnetic correlations up to high temperature
in consequence of the frustrated magnetic structure of
NiS$_2$ \cite{rf:00matsuura}.
In addition, the structural instability
pertinent to this compound was clearly observed both in dilatometric
\cite{Naga76,Miy92} and
$X$-ray diffraction \cite{Jarr73,Thio95} measurements.

Extensive investigations for a series of transition metal dichalcogenides,
MX$_{2}$, during the '70s \cite{wilson85}
appeared to conclude that a simple view of the narrow band concept
applied for the {\it d} electrons on M sites is sufficient to explain
most of the experimental facts.
However, the
recent photoemission experiments indicate that
NiS$_{2}$ is not even a simple Mott Hubbard insulator
but rather a charge transfer insulator (CTI) \cite{fujimori96}.
Note that such CTIs are commonly realized in the strongly correlated
electron systems
of the progenitors of high $T_{C}$ superconductor and colossal
magneto-resistance materials.

Experimental study was done using 9 single crystals
grown by the chemical vapor transport method,
with a total volume of 0.9 cm$^{3}$, which were aligned in the ({\it hkk})
reciprocal lattice zone using shiny (100) and (111) facets.
Neutron scattering measurements were performed on 
TOPAN spectrometer installed at the JRR-3M
Reactor of the Japan Atomic Energy Research Institute.
The crystal structure was determined to be of high symmetry (Pa3)
above $T_{N2}$.
Although a small distortion occurs below $T_{N2}$,
it is ignored in our notation of the crystal symmetry presentation
and we treat all the neutron scattering data within the $fcc$ unit
cell notation.

\begin{figure}
\centerline{\epsfxsize=3in\epsfbox{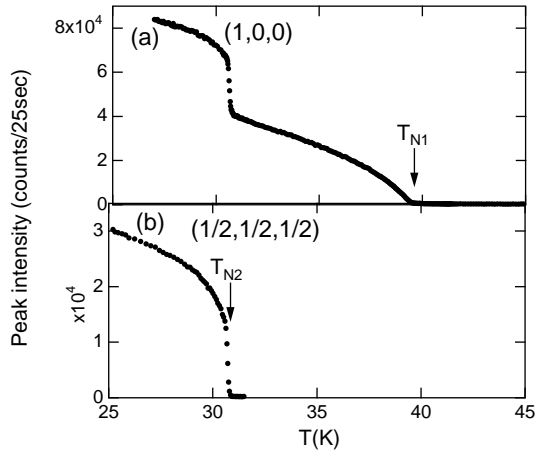}}
\caption{The temperature dependence of the peak intensities of 
the type I (a) AF Bragg (100)
and the type II (b) AF Bragg ($1/2,1/2,1/2$) peaks.}
\label{fig1}
\end{figure}
Elastic neutron scattering measurements confirmed the existence of
\cite{rf:hasting70,Thio95} two types of sharp Bragg reflections:
the type I LRO with AF wave vector
${\bf Q}_{{I}}=(1,0,0)$ appears at $T_{{N1}}=39.6$K and
the type II reflections with ${\bf Q}_{{II}}=(1/2,1/2,1/2)$
arise abruptly below
$T_{{N2}}=30.6$ K superposed on the type I AF LRO with rather
discontinuous phase transition (Fig.~1).
Furthermore,
evaluation of the magnetic entropy estimated from the data of heat
capacity reveals that
the sum of the entropy released
at {\it both} transitions ($\Delta S$ is 0.51 and 0.60 (J/mole$\cdot$K)
for $T_{N1}$ and $T_{N2}$, respectively)
is much smaller than {\it R} ln3 (9.13 J/mole$\cdot$K).
In order to probe spin fluctuations which is characteristic of these
unusual properties of the phase transitions,
we have performed inelastic neutron scattering measurements
with momentum transfers in the $(h00)-(0kk)$ plane of reciprocal space
where both the type I and the type II AF Bragg reflections
are observable.

\begin{table}
\centerline{\epsfxsize=3.2in\epsfbox{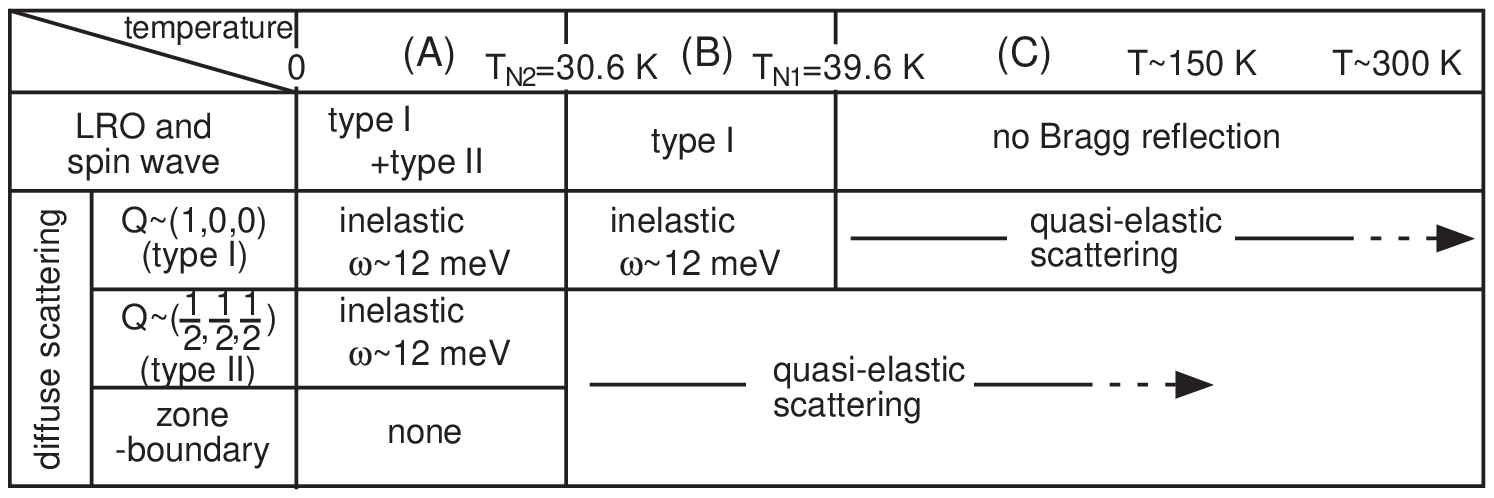}}
\caption{Three characteristic temperature regimes of magnetic excitations in
NiS$_{2}$.}
\label{table1}
\end{table}
Spin fluctuations in NiS$_{2}$ were found to be dramatically
different in three temperature regions as shown in Table~1.
(A) In the first regime of the paramagnetic phase above $T_{N1}$,
strong antiferromagnetic spin correlations
persist up to high temperatures.
Spin fluctuations in this phase are manifested by two components:
(i) magnetic diffuse scattering around the type I AF Bragg point
and (ii) magnetic diffuse scattering component along the {\it fcc}
zone-boundary  around the type II AF Bragg point.
Both components are observable even at room temperature
but grow below 150K,
where anomalies in the conductivity appear.
(B) In the second, intermediate regime $T_{{N2}}<T<T_{{N1}}$,
type I AF LRO arises and concomitantly spin-wave excitations occur.
Two diffusive magnetic scattering components still exist as well.
However, the diffuse scattering around the type I AF Bragg point
becomes rather inelastic as described below.
(C) In the low temperature regime below $T_{N2}$,
where both the type I and II AF LROs are present with a
possible rhombohedral lattice distortion,
the spin waves for each AF LRO were observed.
On the other hand, the diffuse scattering around both
the type I and II AF Bragg points becomes inelastic
and the zone boundary component disappears \cite{unpublished}.

Next, we present the detailed experimental results
for the intermediate regime B followed by theoretical analysis.
At $T=32$ K, we measured
the ${\bf Q}$-scans for the fixed energy transfers of $\omega=2$ meV
and 10 meV along the [100] direction around the
(0, $1+\zeta$, $1+\zeta$) points with $\zeta=0, 0.125, 0.25, 0.375, 0.5$
(Fig.~2a).
For energy transfer $\omega=2$ meV (open circles),
a well defined single peak was observed for $0\leq\zeta \leq0.25$,
whereas it splits into two distinct peaks for $0.25<\zeta \leq0.5$.
Note that the peak for $\zeta=0$ coincides with the type I AF Bragg
point $(0,1,1)$ and that both $(\pm 1/2,3/2,3/2)$
peaks are at the type II AF Bragg points.
These peaks are located on
the zone-boundary line of the $fcc$ lattice
as shown by broken line forming a  honeycomb pattern in Fig.~2a.
The same set of scans with $\omega=10$ meV (closed circle) reveal
basically the same geometric picture,
although the signal intensity
is more concentrated around the type I AF Bragg point.
(We define this as a type I peak.)
The neutron scattering signal is purely magnetic, since the
scattering intensities at momentum transfer
(300) follows the square of the Ni$^{2+}$ ion magnetic
form factor.

We observed that the type I peak
persists even at $\omega=16$ meV
with a frequency-independent peak width of 0.11 \AA$^{-1}$.
Based on this observation,
we conclude that this mode is intrinsically inelastic 
at the intermediate temperature regime (B).
This conclusion is supported by
the energy-scans with fixed {\bf Q} at the type I AF
Bragg point (011) in Fig.~2b.
Furthermore,  {\bf Q}-scans at the type I point 
along [100] and [011] directions
show that the {\it magnetic scattering with $\omega>10$ meV around the type I
Bragg point exhibits an isotropic peak in the (h00)-(0kk) plane}.
We then observed a lower energy diffuse component
around the type II AF Bragg point,
which has a much stronger intensity than the type I peak.
The {\bf q} width and scattering intensity of this component
does not change with {\bf q} along the line
connecting the (0,1.25,1.25) and (0.5,1.5,1.5) points
within the limit of experimental errors.
Therefore, we conclude that {\it the magnetic scattering
around the type II Bragg point is extended on the $fcc$ zone boundary}.
This component was concluded to be quasi-elastic by confirming
energy scan at (1, -0.3, -0.3)
which is close to the zone-boundary line.
\begin{figure}
\centerline{\epsfxsize=3.2in\epsfbox{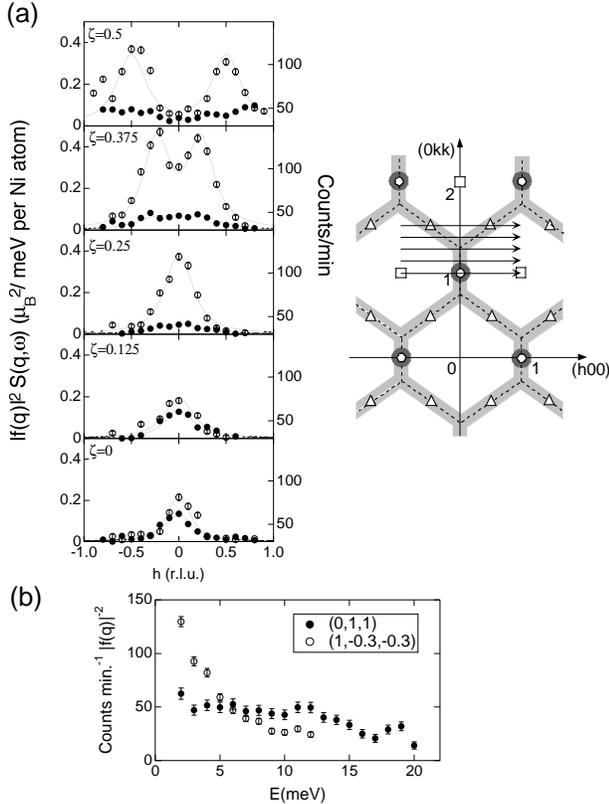}}
\caption{(a) Constant-energy scans along the [100] direction at 32 K.
Open and closed circles represent the data for 
$\hbar\omega=2$ and 10 meV, respectively.
The horizontal collimator is B-30'-60'-B with $E_{{f}}=14.7$ meV.
Absolute units for S({\bf q},$\omega$) were calibrated 
using the intensity of the transverse acoustic phonon.
The right panel shows Reciprocal space for the $fcc$ lattice.
The broken line shows the honeycomb zone boundary 
of the {\it fcc} lattice in the (h00)-(0kk) plane.
Open circles, triangles and gray rectangles represent the AF type
I, type II, and nuclear Bragg points, respectively.\\
(b) Constant-{\bf Q} scans at the type I AF Bragg point (011) 
and near the {\it fcc} zone boundary (1,-0.3,-0.3) at 32 K. 
Open and closed circles show the
data at (011) and (1,-0.3,-0.3), respectively. 
The data are divided by the square of the magnetic form factor 
at each position after subtraction of the background.}
\label{fig2}
\end{figure}

To extract information from the observed geometrical pattern of the
critical scattering,
we studied the momentum-transfer dependence of the
energy integrated scattering law $S({\bf Q})=\int d \omega
S({\bf Q},\omega)/ \mid F({\bf Q}) \mid^2$
( $F({\bf Q})$ is the magnetic form factor),
which is proportional to the real part of
the static momentum dependent magnetic susceptibility.
Both above and below the N\'eel temperature
$T_{N1}$ the  static susceptibility 
can be expressed in following form, 
$
\chi'({\bf Q}) \sim
\left\{ \delta+(1-{\cal L}({\bf Q})/{\cal L}({\bf Q_{AF}}) \right\}^{-1}
$.
The parameter $\delta$ specifies how much the actual temperature 
deviates from $T_{N1}$,
${\bf Q_{AF}}$ is an AF wave vector 
where the divergence of susceptibility at $T=T_{N1}$ occurs,
and ${\cal L}({\bf Q})$ represents 
the Fourier transform of the interaction matrix.
Considering the NN $J_{1}$, next NN
$J_{2}$ and next next (third) NN $J_{3}$ AF interactions,
one can express the Fourier transform of the interaction matrix
$
{\cal L}({\bf Q}) = - \mid J_{1} \mid
\left\{
\phi_{1}({\bf Q}) +
R_{2} \phi_{2}({\bf Q}) +
R_{3} \phi_{3}({\bf Q})
\right\}
$
in terms of the standard $fcc$ lattice geometrical factors $\phi_i$ and
ratios
$R_{2}=J_{2}/J_{1}$ and $R_{3}=J_{3}/J_{1}$.
\begin{figure}
\centerline{\epsfxsize=3.2in\epsfbox{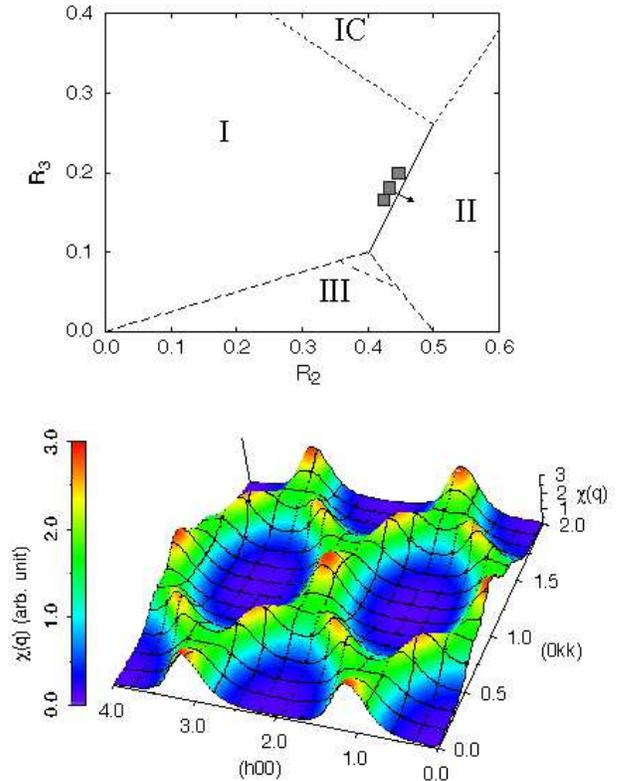}}
\caption{Upper panel: Magnetic phase diagram of the $fcc$ lattice.
Type I, type II and type III (${\bf Q}_{AF}=(1,1/2,0)$) phases 
are indicated by I, II, and III, respectively.
The area denoted by IC and the small triangle near the crossing
of the lines separating I--II and II--III phases are regions where
incommensurate phases are found.
Gray rectangles are possible locations
of NiS$_2$ in the phase diagram in the high symmetry phase.
The arrow represents the possible change of the position 
in the diagram due to rhombohedral structural distortion.\\
Lower panel: Surface of critical scattering intensity for $R_2=0.434$,
$R_3=0.164$, and $\delta=0.3$.}
\label{fig3}
\end{figure}

Since the value of the AF wave vector depends 
only on the parameters $R_2$ and $R_3$,
we analyzed the phase diagram in terms of these coordinates
(Fig.~3, upper panel).
We found that the honeycomb structure (Fig.~3, lower panel)
of the critical scattering in the $(h00)-(0kk)$ plane is
{\it a general feature of the $fcc$ lattice with the AF interactions
up to third NN}.
This pattern is robust in a rather large domain where $R_2<0.6$ and
$R_3<0.3$.
Therefore, to locate the position of NiS$_2$ in the phase diagram,
one have to study the fine features of the honeycomb structure
that are general for the $fcc$ lattice .
To fix the values $R_2$ and $R_3$
for NiS$_2$ in the high-symmetry phase,
i.e. for $T>T_{{N2}}$,
we calculated the critical scattering pattern
at numerous points in phase I,
and found three points (gray rectangles in upper panel of Fig.~3)
that obey the following conditions obtained experimentally:
(i) the scattering around the type I AF Bragg point is a symmetric
circle-like peak;
(ii) the contour plot of the scattering intensities
around the type II AF Bragg points are  ellipses aligned along the $fcc$
zone boundary; and
(iii) energy integrated intensities at the type I and 
the type II AF Bragg points
are equal within the limit of experimental errors.
Therefore,
one can conclude that NiS$_2$ in a high-symmetry phase
is situated in the vicinity of the gray rectangles and,
hence, very close to the border separating
the I and II phases (see Fig.~3) \cite{SOL}.
This quasi-degeneracy also explains the
anomalous temperature dependence of the spin fluctuations 
above $T_{N1}$ (regime A),
namely the type I AF LRO temperature is
suppressed by the frustrated magnetism,
and the magnetic diffuse scattering is
expected for a wide temperature range above $T_{N1}$.

Here, we discuss the fact that the two components
of magnetic scattering have different properties
in the intermediate temperature regime (B).
These components belong to different irreducible representations
of the symmetry group of the cubic lattice.

Even when the type I AF LRO is established,
both components are still present.
Hence, the virtual ground state of the type II 
caused by magnetic frustration is
quasidegenerate with the  actual type I magnetic ground state
of the cubic lattice.
We emphasize that the quasidegeneracy occurs only accidentally between
the two ground states which is incompatible in the cubic lattice
symmetry (i.e. can be removed by a change of AF couplings)
according to the Wigner theorem in group theory.
Therefore, the system is not exactly frustrated but nearly frustrated.
However, the nearly frustrated situation can be resolved by the structural
distortion which lowers the lattice symmetry.
As a consequence, the type I and the type II AF LROs become compatible 
in the reduced symmetry group.
In order to elucidate possible consequences of SPT
we studied small deviations of the magnetic exchange parameters
constrained by the symmetry of the rhombohedral phase.
Then we found that comparatively small changes of
magnetic couplings ($<10^{-2}$) can induce a shift in the phase diagram
denoted by the arrow in Fig.~4.
A notable feature of the lowest transition at $T_{N2}$ is that
the {\it system avoids the phase of nearly frustrated magnetic
interactions by lowering the lattice symmetry}.

In conclusion, in the paramagnetic and high symmetry phase (A) above
$T_{N1}$,
NiS$_2$ is a paramagnet where anomalies are governed
by geometrical frustration and competing interactions.
The second-order transition at $T_{N1}$ into the LRO type I AF state
does not remove the nearly frustrated situation.
The magnetic diffuse scattering along the $fcc$ zone boundary,
indicating the quasidegeneracy of the type I and II AF LROs,
persists down to the frustration-avoiding SPT at $T_{N2}$,
where the type I and the type II phases probably become
compatible due to the rhombohedral structural distortion.

Authors are indebted to S. Maekawa, W. Koshibae and D. Belanger for
stimulated discussions.
M. Matsuura has been supported by Research Fellowships of the Japan Society
for the Promotion of Science for Young Scientists.


\begin{thebibliography}{99}
\bibitem{Ra01}
For a review see A.\ P.\ Ramirez, in ``Magnetic Materials''
(North-Holland, Amsterdam, to be published).
%
\bibitem{Lee00}
S.-H.\ Lee {\it et al.}, Phys.\ Rev.\ Lett.\ {\bf 84}, 3718 (2000).
%
\bibitem{Melzi00}
R.\ Melzi {\it et al.}, Phys.\ Rev.\ Lett.\ {\bf 85}, 1318 (2000).
%
\bibitem{rf:yao96}
X.\ Yao {\it et al.}, Phys.\ Rev.\ B {\bf 54}, 17469 (1996).
%
\bibitem{Yao97}
X.\ Yao {\it et al.}, Phys.\ Rev.\ B {\bf 56},  7129 (1997).
%
\bibitem{rf:00matsuura}
M.\ Matsuura {\it et al.}, J.\ Phys.\ Soc.\ Japan {\bf 69}, 1503 (2000).
%
\bibitem{rf:hasting70}
J.\ M.\ Hasting, L.\ M.\ Corliss, IBM J.\ Res.\ Dev.\ {\bf 14}, 227 (1970).
%
\bibitem{rf:78kikuchi}
K.\ Kikuchi {\it et al.}, J.\ Phys.\ Soc.\ Japan {\bf 45}, 444 (1978).
%
\bibitem{Naga76}
H.\ Nagata, H.\ Ito, T.\ Miyadai,
J.\ Phys.\ Soc.\ Japan {\bf 41}, 2133 (1976).
%
\bibitem{Miy92}
T. Miyadai, M.\ Saitoh, Y.\ Tazuke,
J.\ Magn.\ Magn.\ Mat.\ {\bf 104-107}, 1953 (1992).
%
\bibitem{Jarr73}
H.\ S.\ Jarrett {\it et al.}, Mat.\ Res.\ Bull.\ {\bf 8}, 877 (1973).
%
\bibitem{Thio95}
T.\ Thio, J.\ W.\ Bennett, T.\ R.\ Thurston,
Phys.\ Rev.\ B {\bf 52}, 3555 (1995).
%
\bibitem{wilson85}
J.\ A.\ Wilson, {\it The metallic and non-metallic states of matter}
(Taylor \& Francis, 1985) Chap.\ 9.
%
\bibitem{fujimori96}
A.\ Fujimori {\it et al.}, Phys.\ Rev.\ B {\bf 54}, 16329 (1996).
%
\bibitem{unpublished}
T.\ Thurston {\it et al.}, unpublished data.
%
\bibitem{SOL}
The LDA + U calculation gives values of $J_1$, $J_2$, $J_3$
which are consistent with this estimation in upper panel of Fig.~3;
I.\ V.\ Solovyev, A.\ Mishchenko, N.\ Nagaosa, in preparation.
%
\end{thebibliography}
\end{document}